# Sustaining the Montage Image Mosaic Engine Since 2002

G. Bruce Berriman*[a], John C. Good[a]
[a] Caltech/IPAC-NExScI, 1201 East California Blvd., Pasadena, CA 91125, USA

## ABSTRACT

This paper describes how we have sustained the Montage image mosaic engine (http://montage.ipac.caltech.edu) first released in 2002, to support the ever-growing scale and complexity of modern data sets. The key to its longevity has been its design as a toolkit written in ANSI-C, with each tool performing one distinct task, for easy integration into scripts, pipelines and workflows. The same code base now supports Windows, JavaScript and Python by taking advantage of recent advances in compilers. The design has led to applicability of Montage far beyond what was anticipated when Montage was first built, such as supporting observation planning for the JWST. Moreover, Montage is highly scalable and is in wide use within the IT community to develop advanced, fault-tolerant cyber-infrastructure, such as job schedulers for grids, workflow orchestration, and restructuring techniques for processing complex workflows and pipelines.

## 1. INTRODUCTION

The Montage image mosaic engine is entering its sixteenth year of support to the astrophysics and IT communities. The mosaics created by Montage are intended for scientific analysis: they preserve the calibration and astrometric fidelity of the original images, and rectify the highly variable sky background to a common level across all the original images[1]. Montage has found applicability far beyond its original goal of enabling scientific analysis of images acquired in wide-area astronomy surveys. It has been applied to, among other things, searches for near-Earth objects[2], observation planning for JWST[3], studies of Raman scattering in laser AO systems[4], development of products for citizen science projects[5,6], and creating data products used by machine-learning algorithms for interpreting photometric redshifts[7]. The IT community has adopted it as an exemplar application in developing advances in cyber-infrastructure[8]. This breadth of use has come about because Montage creates science-ready images; because of its design as a portable toolkit, which offers flexibility to end users and which enables integration into workflows; and because it has responded to the dramatic changes in the data and computational landscapes in astronomy[9], including multi-dimensional image data sets (hereafter, "data cubes") and the Hierarchical Equal Area isoLatitude Pixelization (HEALPix) sky tessellation scheme now standard in cosmic background experiments[10].

The purpose of this paper is to describe in detail how the design of Montage promotes sustainability, to explain how the astronomical and IT communities have exploited the design, and to explain how a user community was built around Montage. The narrative includes aspects of the "back story" behind the design decisions and community involvement. It emphasizes the value of design in promoting sustainability, the use of new technologies where they provide value, the need to talk to the user community, and how to rectify decisions that, in hindsight, proved incorrect. The code is freely available from GitHub (https://github.com/Caltech-IPAC/Montage) or the Montage website (http://montage.ipac.caltech.edu/docs/download.html).

## 2. DESIGN FOR SUSTAINABILITY

### 2.1 Architectural Approach

By design, Montage was written as an ANSI-C toolkit that creates mosaics in response to the user's specifications of output coordinate system, image reprojection, pixel sampling, and image rotation angle. The toolkit contains components that perform the tasks needed to create such mosaics: reprojection and resampling of the input images; rectification of the variable sky and instrumental background across the images to a common level; and co-addition of the reprojected and rectified images. It also contains utilities for performing tasks such as managing large-scale mosaics and analyzing the metadata of FITS files for content and completeness[1].

Each component performs one task in the calculation of a mosaic. These tasks are as follows:

- Analyze the geometry of input images to derive geometry of output mosaic.
- Re-project the input images.
- Model background images and rectify to a common level.
- Co-add the processed images.

The schematic shown in Figure 1 shows how the components operate in an end-to-end processing flow.

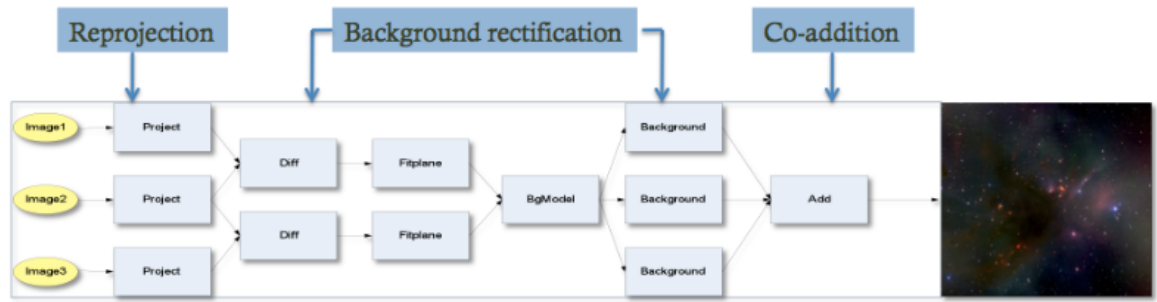

Figure 1. End-to-end processing flow in the creation of an image mosaic with the Montage imaged mosaic engine.

The processing flows scale from desktops, where they are usually run serially through scripts, to high-performance platforms, where they are parallelized through workflow managers such as Pegasus[8], or through the Message Passing Interface (MPI).

**2.2 Design Drivers**

A brief history of the early science drivers in 2002 will provide context for the toolkit design. When the project began, there were two science drivers guiding the development of Montage. One was to create mosaics from the 2 million individual survey images acquired by the Two Micron All Sky Survey (2MASS) over the full-sky in the J (1.25 μm), H (1.65 μm), and $K_s$ (2.2 μm) bands[11]. These mosaics would enable study of the many astrophysical structures larger than the individual images, which are 512 by 1024 pixels (1"/pix) in size, if they preserved the calibration and astrometric fidelity of the input images. The second driver was to combine the 2MASS images with images from other missions and telescopes to create multi-color images that had the same image parameters as the 2MASS images.

The 2MASS images contain a very bright sky background that varies across each image, and from image to image. Creation of mosaics required rectification of this background emission to a common level across each image, and these characteristics mandated a more sophisticated scheme than matching background levels at the edges of images or artificially flattening the background levels across images. The scheme chosen fitted the differences in background levels across each image to create a model of the sky background. The model values are then subtracted from the pixel value of each image. This background rectification scheme is, to our knowledge, unique to Montage. Creation of multi-color images and preservation of the astrometric and calibration fidelity of the input images were both achieved with a spherical trigonometry algorithm to co-register images on the sky and to compute the flux in the output images.

The final design that responded to the drivers was a toolkit written in ANSI-C rather than an application, with each component performing one task. The reason for this decision was to enable ease of maintenance and simplify testing, and was in fact against the recommendations of a number of scientists. The components represented in Figure 1 are all individual components inside the toolkit.

## 2.3 Design Details

The design and processing flow represented in Figure 1 remain in place today. In addition responding to the science drivers, there was, from the beginning, a conscious decision to design Montage with long-term sustainability and ease of maintenance in mind. It was written in ANSI-C for performance and for portability across all common *nix platforms, avoids commercial third-party dependencies, uses only common, freely available libraries, and does not use shared memory. All modules accept input from the command line and return structured American Standard Code for Information Interchange (ASCII) responses that can be parsed by any computer. The toolkit is self-contained with all necessary support libraries, and built with a `make` command. The libraries include the Smithsonian Astrophysical Observatory (SAO) WCSTools library (hereafter, WCSTools; (http://tdc-www.harvard.edu/wcstools/), which implements the World Coordinate System (WCS) transformations between pixels and spatial coordinates of images. By default, the spherical trigonometry algorithm in Montage is able to process all spherical image projections that are supported by WCSTools.

## 2.4 Release History and Support for New Functionality

The "close-to-the-metal" design has enabled Montage to adapt to new technology, and without this design, it is unlikely the functionality incorporated into Montage since 2002 would have been delivered with available resources. The release history is shown in Table 1. Versions 1 to 2 (2003–2005), funded by NASA, offered creation of two-dimensional image mosaics; these early releases contained essentially the modules depicted in Figure 1. Version 3 (2005-2013) included two new groups of modules. One set contained utilities for the organization and management of data, and analysis and repair of image metadata. The second set contained modules to determine sky coverage data from major astronomy image surveys such as: 2MASS, the Digital Sky Survey (DSS), the Sloan Digital Sky Survey (SDSS) and the Wide-field Infrared Survey Explorer (WISE).

Since 2014, now with NSF support, releases have responded to the evolution of the astronomy data landscape. Version 4 offered support for data cubes. Rather than make existing reprojection modules more complex, Version 4 took the approach of adding new modules for re-projecting and sub-setting data cubes. Version 5 (2016) supports processing of HEALPix data, now the standard scheme for organizing wide-area cosmic-background data sets, and the Tessellated Octahedral Adaptive Subdivision Transform (TOAST)[12], required for the consumption of images in the World Wide Telescope. Montage takes the approach of treating these two sky-tessellation schemes as WCS projections so that all the functionality in Montage is automatically accessible to them[13]. This is achieved at the expense of installing a custom WCSTools library, as HEALPix and TOAST are not yet supported by official releases of the library. Altogether, there have been over 20,000 downloads to date of the various releases from independent IP addresses.

Table 1. The release history of Montage from 2002 to 2018.

| Release | Date(s) | Content | License | Funding |
|---|---|---|---|---|
| 1.0 -2.2 | 2002-2005 | Core modules for building mosaics | Proprietary | NASA |
| 3.0-3.3 | 2006-2013 | Utilities, data access | Proprietary | … |
| 4.0-4.1 | 2015 | Data cubes | BSD 3-clause | NSF |
| 5.0 | 2016 | HEALPix, TOAST, fast-reprojection algorithm. Build as library or toolkit. | BSD 3-clause | NSF |
| 6.0 | 2018 (scheduled) | Windows | BSD 3-clause | NSF |
| 7.0 | 2018 (scheduled) | Build as Python library | BSD 3-clause | NSF |

## 2.5 Windows and JavaScript

Versions 1 through 5 ran on all common variants of *nix. We have taken advantage of advances in compiler technology to compile the same code base that runs under *nix to run under Windows 10 and JavaScript. While Montage can be distributed across platforms via container technologies such as Docker, there is in fact no necessity to do this. The Minimalist GNU for Windows (`MinGW`) compiler includes ports of the GNU compiler, linker and libraries that run on Linux and create output executable images that run on Windows machines. `MinGW` creates Windows executable images for all but the data access modules that used secure sockets, supported on Linux but not Windows. This release, listed as Version 6.0 in Table 1, is scheduled for deployment in summer 2018 following completion of evaluation by members of the E/PO and amateur astronomy communities.

Modern compilers first build a Low Level Virtual Machine (LLVM) "intermediate representation" of an executable image. The `Emscripten` tool interprets this LLVM to build executable code that runs under JavaScript in a browser at approximately 50% of compiled speeds. E. Mandel has taken advantage of this to integrate Montage into the JS9 environment (https://js9.si.edu), a web version of the DS9 desktop application that supports astronomical image display to web, desktop, and mobile environments. Montage is primarily used by JS9 to create co-registered images at multiple wavelengths that are then blended for display. Figure 2 shows an example of such images of the colliding galaxies NGC 2027 and IC 2163.

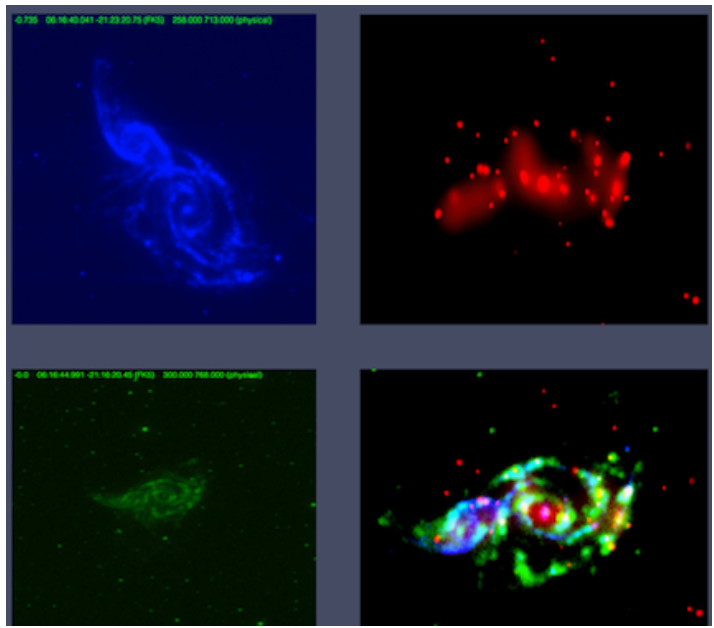

Figure 2: Montage JavaScript extension modules are used by JS9 web-based image display (https://js9.si.edu) to perform flux-conserved WCS reprojections. The resulting aligned images can then, for example, be blended together to facilitate feature identification and analysis. The figure shows an example: images of the colliding galaxies NGC 2027 and IC 2163 from Chandra (red), Spitzer (blue), and Galex (green) resampled and re-projected in JS9 to produce a blended 3-color image (Image courtesy of E. Mandel).

### 2.6 Python

A transformation in astronomy science environments is taking place, driven by the adoption of Python as the programming language of choice in astronomy: 67% of astronomers surveyed in 2015[14] now use Python. Python is best thought of as a framework that enables codes to be exposed in a common environment and astronomers are taking advantage of it to perform complex analyses that once required manual effort to locate, build, apply and integrate software packages. In particular, linking C/C++ libraries to the Python kernel at run time as binary extensions allows them to run under Python *at compiled speeds*. Libraries used every day in Python—such as NumPy for numerical calculation and Matplotlib for visualization—are of this type. We have therefore prototyped building Python binary extensions of the "core" Montage modules required to create a mosaic. Transforming a Montage module to a Python binary extension involves turning the code into a C library, with driver code fully separated to reproduce the calling sequence of the command-line tools; and then adding Python and C linkage code with the Cython library, which acts as a bridge between general C libraries and the Python interface.

As a proof-of-concept, we have built the Montage core modules for Python 2.x and Python 3.x on Mac OS X and on all common Linux platforms. The build on Macs is straightforward as it is a uniform platform, but this is not the case for Linux. A uniform build in this case was achieved by compiling with a Docker container built for CentOS 5.11, which ensures consistent use of system-level functionality across all flavors of Linux. Inside Python, the Montage library is installed simply through the `pip` Python installer. No other Python modules or OS-level installations are required. We describe two examples of the use of Montage running under Python.

**Prototype One: Creation of a mosaic in a Jupyter Notebook.** Figure 3 shows a sample of a section of a Notebook developed to create a mosaic of 2MASS images (the whole Notebook is too large to display in this paper). The Figure shows the reprojection and background rectification steps; the full steps in the complete Notebook are identical to those used in *nix scripts. This prototype ran to completion with the same execution time as a Linux script and creates the mosaic in Figure 4.

```
In [18]:  # Reproject the original images to the  frame of the
          # output FITS header we created

          rtn = mProjExec("raw", "rimages.tbl", "region.hdr", projdir="projected", quickMode=True)

          print("mProjExec:         " + str(rtn), flush=True)

          mImgtbl("projected", "pimages.tbl")

          print("mImgtbl (projected): " + str(rtn), flush=True)

          mProjExec:           {'status': '0', 'count': 49, 'failed': 0, 'nooverlap': 0}
          mImgtbl (projected): {'status': '0', 'count': 49, 'failed': 0, 'nooverlap': 0}
```

```
In [22]:  # Background correct the projected images.

          rtn = mBgExec("projected", "pimages.tbl", "corrections.tbl", "corrected")

          print("mBgExec:           " + str(rtn), flush=True)

          rtn = mImgtbl("corrected", "cimages.tbl")

          print("mImgtbl (corrected): " + str(rtn), flush=True)

          # Coadd the background-corrected, projected images.

          rtn = mAdd("corrected", "cimages.tbl", "region.hdr", "mosaic.fits")

          print("mAdd:              " + str(rtn), flush=True)

          mBgExec:             {'status': '0', 'count': 49, 'nocorrection': 0, 'failed': 0}
          mImgtbl (corrected): {'status': '0', 'count': 49, 'badfits': 0, 'badwcs': 0}
          mAdd:                {'status': '0', 'time': 1.0}
```

Figure 3: Sections of a Jupiter Notebook showing two of the steps in creating a mosaic of M17 with 2MASS data in the K band (2.2 μm): reprojection of images (top), and background rectification (bottom).

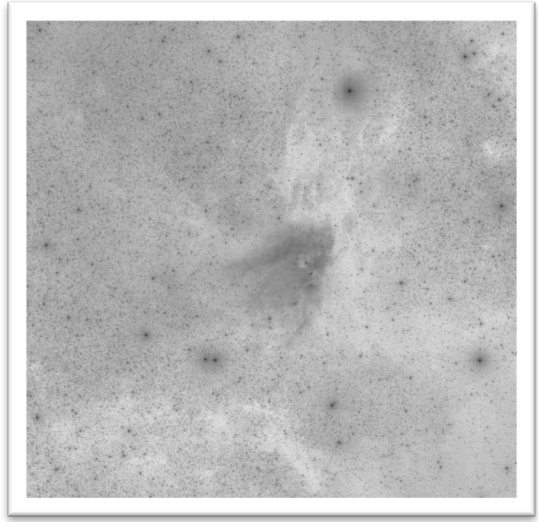

Figure 4: Mosaic of M17 created with the Notebook in Figure 3.

**Prototype Two:** An interface to the Montage Graphics Engine. The second example is a graphical interface to mViewer, created with PyQt, a Python binding to the cross-platform Graphical User Interface (GUI) toolkit Qt. Figure 5 shows a screenshot of the viewer control panel in the GUI, and an image of the sky-coverage of observations of M51 created with it.

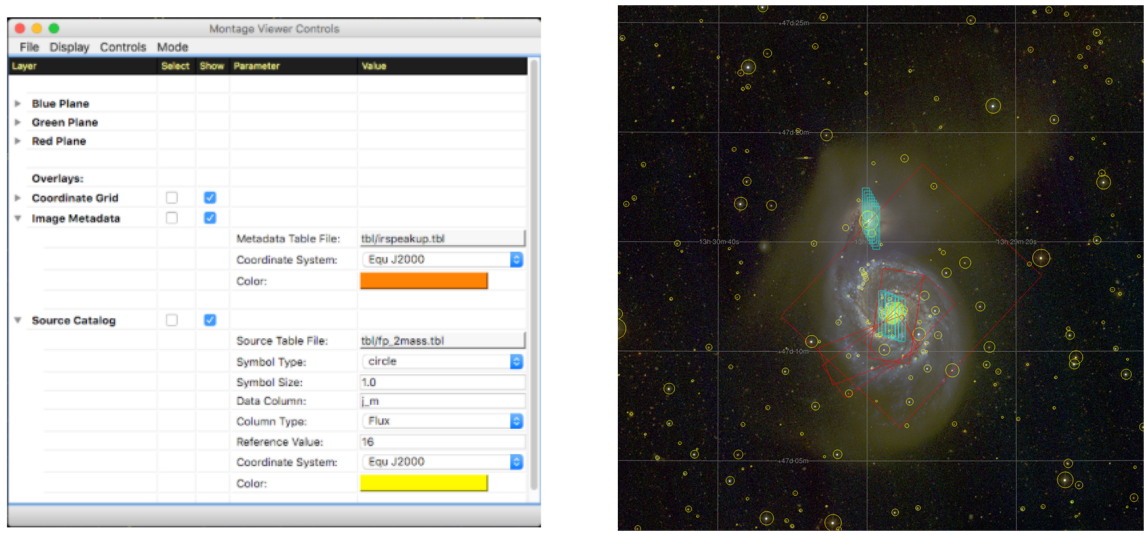

Figure 5: Image of M51 (right) created in JupyterLabs with the graphics engine and GUI (left) in Montage. Key: Boxes - Red: Spitzer/IRAC, Green: Spitzer/MIP, Yellow: HST/WFPC, Blue: HST/WFPC2, Purple: HST/NICMOS. Circles- Yellow: 2MASS point sources.

## 3. APPLICATIONS OF MONTAGE BY THE COMMUNITY

### 3.1 Integration into Processing Environments, Workflows and Pipelines.

The individual modules are designed as useful tools in their own right. Astronomers have used the Montage background rectification modules in preparing images for research[15,16,17]. Data providers can integrate the components into their processing environments, workflows and pipelines. Among the earliest adopters of this approach were the Spitzer Space Telescope Legacy Projects, selected because of their exceptional importance to astronomy and guaranteed telescope time in return for early public release of data. The Spitzer Infrared Nearby Galaxies Survey (SINGS) used Montage to create multi-images of the survey galaxies. Surveying the Agents of a Galaxy's Evolution (SAGE), a 24- and 70-Micron Survey of the Inner Galactic Disk with MIPS (MIPSGAL), the Spitzer Wide-area InfraRed Extragalactic Survey (SWIRE), and the Galactic Legacy Infrared Midplane Survey Extraordinaire (GLIMPSE; actually a group of related Galactic Plane surveys) integrated Montage into their data production environments. GLIMPSE360, which aggregated the GLIMPSE data sets into a single multi-color map of the Galactic Plane, used Montage as a reprojection engine[18]. These projects themselves have high impact, with over 8,000 cumulative citations to date (through searches made by the authors of this paper in ADS Labs). To our knowledge, 27 projects have, to date, used Montage in this way. Recent ones and their citation statistics (also derived by the current authors) are shown in Table 2.

Table 2: Recent projects integrating Montage into their data processing environments, and their citation rates to April 2018.

| Mission, Project or Service | Citations |
| --- | --- |
| AKARI | 501 |
| Australian Square Kilometre Array Pathfinder (ASKAP) | 301 |
| INT Photometric H-Alpha Survey (IPHAS) | 532 |
| SuperCOSMOS | 285 |
| Bolocam Galactic Plane Survey (BGPS) | 242 |
| Cosmic Background Imager (CBI) | 192 |
| SAGES Legacy Unifying Globulars and GalaxieS (SLUGGS) | 119 |
| Astronomical Plotting Library in Python (APLpy) | 65 |

### 3.2 Montage as a Visualization Engine

Montage is finding growing applicability as a graphics and visualization engine[13] primarily because it responds to two Computational Grand Challenges[19]: automated creation of images from large collections of data; and integration of visualization tools into pipelines and workflows. The JS9 images shown in Figure 2 are an example of integration into a visualization system. Montage has been integrated into a client-server architecture intended as a demonstration of how a visualization environment would operate when extended to petascale processing[20]. A further example[21] describes how cutout images can be incorporated into the Virtual Reality (VR) workflows for use in commodity head-mounted virtual reality displays.

Version 4 includes a new graphics engine, `mViewer`, which includes an adaptive stretch algorithm. It can create multi-color visualizations, and can also be used as a sky graphics engine. Figure 4 shows an example of the kind of the latter: a sky coverage map of Exoplanet projects as part of the preparatory work for the Transiting Exoplanet Survey Satellite (TESS)[22].

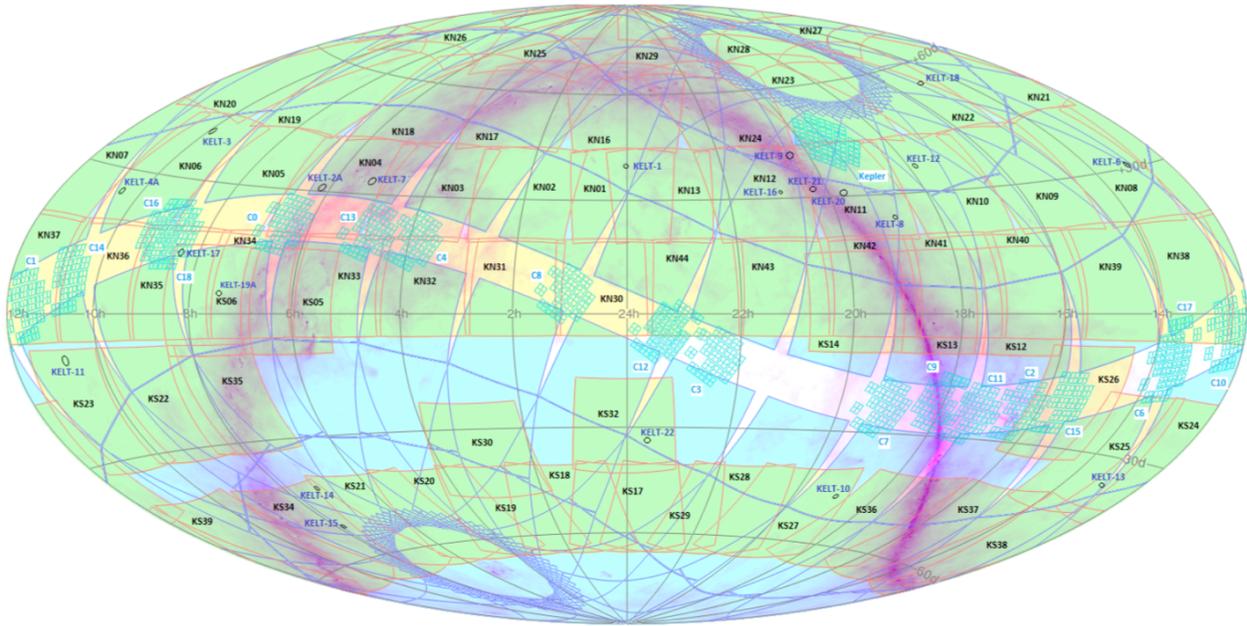

Figure 6: Sky coverage maps of the Kilodegree Extremely Little Telescope (KELT) and Transiting Exoplanet Survey Satellite (TESS) fields, created with the Montage graphics engine mViewer22. The KELT fields are outlined in orange and representative TESS fields are outlined in purple. Regions where KELT fields overlap TESS fields are green. Regions where TESS fields have no KELT field overlap are light blue. Regions with no TESS coverage are yellow. The KELT-North fields are labeled KN01 - KN44 and the KELT-South fields are labeled KS05 - KS36. Also shown are the Kepler and K2 fields outlined in cyan, a model of the galactic plane in magenta, and the locations of the planets discovered by KELT identified by dark blue labels. (Image courtesy of J. Pepper.)

### 3.3 Adoption by IT Community

In 2002, the Pegasus Workflow Manager used Montage as one of the driver applications to optimize its design. Since then, it has become an exemplar application in developing advanced infrastructure, and one-third of the tickets submitted to the Help Desk now come from the IT community. It has been widely used in the following areas, where has been cited 486 times in the IT literature since 2014:

- Task scheduling in distributed environments (performance-focused).
- Designing job schedulers for the grid.
- Designing fault tolerance techniques for job schedulers.
- Exploring issues of data provenance in scientific workflows.
- Exploring the cost and performance of scientific applications running on Clouds.
- Developing high-performance workflow restructuring techniques.
- Developing application performance frameworks.
- Developing workflow orchestration techniques.
- Transparency in Research and Software Sustainability.

## 4. BUILDING A USER COMMUNITY

Despite the successes described in earlier sections, adoption by the astronomical community as a tool for research on desktops was in fact slow. Even though Montage was released in 2002, it was not until 2005 that citations began to appear in peer reviewed astronomy journals, as shown in the cumulative peer-reviewed citations in astronomy journals from 2005 to date in Figure 6. And yet, high-impact projects and the IT community exploited Montage from the first release. Why was this? There were two reasons, discovered primarily by canvassing astronomers at topical meetings: the toolkit design itself, and performance.

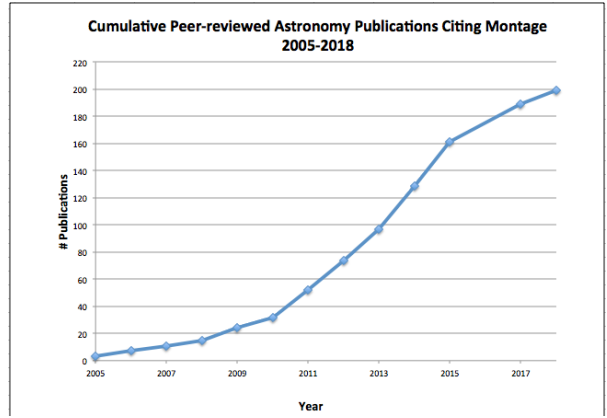

Figure 7: Cumulative number of citations to Montage (2005–2018 March 20) in Ap J, A & A, MNRAS, AJ and PASP.
Source: Search on Montage in the Bumblebee interface of ADS Labs.

Users of early releases expected to find simple scripts that would accept as input the parameters of their mosaics and the location of their data sets. Users were in fact to expected write the scripts needed to create a mosaic and were unprepared to do so, even though the scripts were straightforward.

An example is shown in Figure 8, used to create a three-color mosaic of the Pleiades shown underneath the script. This script was written in 2008 and executed under version 2.2, and it still runs under the latest releases.

```
#!/bin/bash
# Pleiades Image creation BASH script.
# Inseok Song, 2007    for bands in DSS2B DSS2R DSS2IR; do echo Processing
${bands};
mkdir $bands;
cd $bands;
mkdir raw projected;
cd raw;
mArchiveList dss ${bands} "56.5 23.75" 3 3 remote.tbl;
mArchiveExec remote.tbl;
cd .. ;
mImgtbl raw rimages.tbl ;
mProjExec -p raw rimages.tbl ../pleiades.hdr projected stats.tbl ;
mImgtbl projected pimages.tbl ;
mAdd -p projected pimages.tbl ../pleiades.hdr ${bands}.fits ;
cd .. ;
done
mJPEG -blue DSS2B/DSS2B.fits -1s 99.999% gaussian-log \
      -green DSS2R/DSS2R.fits -1s 99.999% gaussian-log \
      -red DSS2IR/DSS2IR.fits -1s 99.999% gaussian-log \
        -out DSS2_BRIR.jpg
```

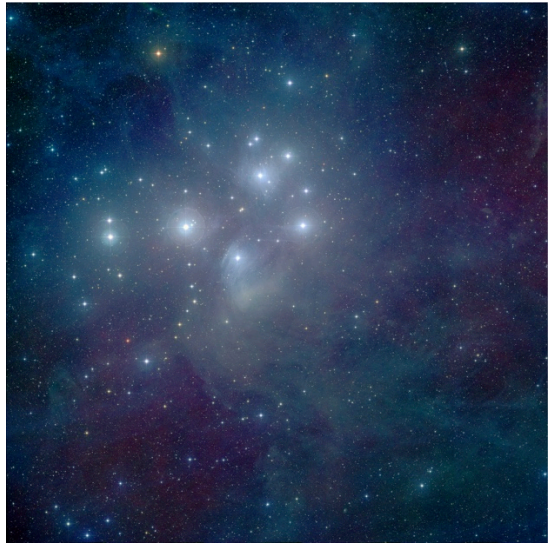

Figure 8: Three-color DSS mosaic of the Pleaides created by Montage by the script included in the text.

Moreover, users were expected to create the FITS header that described the geometry of the image mosaic on the sky. These issues were corrected by providing executive scripts that took care of these tasks, by and publishing tutorials and scripts that users can adapt.  Ironically, the very design that appealed to project developers and the IT community was impeding take-up by astronomers working on their desktops. The second difficulty was that Montage used spherical geometry to reproject the image pattern on the sky from the input to the output image, and the penalty for the generality in this algorithm was slow performance. Re-projection was speeded up by creating a dedicated plane-to-plane algorithm that was applicable to tangent-plane projections, the most common image projection in use on astronomy. The algorithm is an adaptation of an algorithm designed for the Spitzer Space Telescope.

Furthermore, many astronomers expressed the need for fast, occasional access to mosaics they needed occasionally. In response, we released on an-demand web-based service (http://hachi.ipac.caltech.edu:8080/montage). It creates mosaics of up to 1°x 1° in size of four heavily used imaging surveys: 2MASS, SDSS, WISE, and DSS. The service has been in operation since 2008 and has received a total of 37,800 requests from 2,300 registered users for mosaics, and created 1.9 PB of mosaics. The service was in fact an experiment to understand how it could it be built with low-cost hardware. Requests for mosaics are handled by a request management system built for the now-defunct National Virtual Observatory (NVO).

New development is community driven. The principal source of requirements is analysis of tickets submitted to the Help Desk, along with such sources as canvassing users at meetings, responses to the Montage mailing list, and comments on social media, such as the Astronomers Facebook page. A total of 380 requests for support have been answered since 2014. Requests for service upgrades have been incorporated into incremental releases; e.g., support for co-addition of sparsely populated images, requested by the Swift mission; and control of the output data types in image cutouts, requested by ASKAP. Through the Help Desk, astronomers are requesting consultations on increasingly sophisticated topics that include not only questions on Montage, but also on image processing, data formats, and data quality, sky coverage of image surveys, and integration into projects. The Montage Help Desk is becoming more of a resource for the community on imaging questions than simply a Help Desk for Montage itself, and is requiring increasingly complex responses, sometimes requiring tutorial services rather than a simple answer to a question.

## 5. CONCLUSIONS

Montage has been successfully maintained for 16 years and has found applicability far beyond its original goal of creating mosaics of wide-area image mosaics. Changes in the data and computational landscape have been incorporated into the same toolkit design that was deployed in 2002, and the same code base runs on *nix platforms, Mac OS X, Windows, JavaScript and Python 2.x and 3.x. The paper has discussed how the breadth of use of Montage has expanded discussed through sustainable design, careful use of technology advances, response to the evolution of the astronomy data landscape, and the need for persistence in building a user community. The development thus far has used an Open Source, Closed Development model, but the project plans to migrate to an Open Development model in the future.

## 6. ACKNOWLEDGEMENTS

Montage is funded by the National Science Foundation under grant numbers ACI-1440620 and ACI- 1642453, and was previously funded by the National Aeronautics and Space Administration's Earth Science Technology Office, Computation Technologies project, under cooperative agreement number NCC 5-626 between NASA and the California Institute of Technology. We wish to thank our colleagues for participating in the development of Montage: Tom Prince, Daniel S. Katz, Joe Jacob, Anastasia Laity, Roy Williams, Marcy Harbut, Attila Bergou, Angela Lerias, Ewa Deelman, Gideon Juve, and Mats Rynge.